\documentclass[aps,prl,twocolumn,showpacs,floats,amsmath,amssymb,superscriptaddress,groupedaddress,10pt]{revtex4}
\usepackage{epsfig}
\usepackage{subfig}
\usepackage{graphics}
\usepackage{colortbl}
\usepackage{colordvi}
\usepackage{verbatim}
\usepackage{amsmath}
\usepackage{enumerate}
\usepackage{wrapfig}
\usepackage{dcolumn}
\usepackage{bm}

\begin{document}

\title{Criticality in Fiber Bundle Model}

\author{Subhadeep Roy}
\email{sroy@imsc.res.in}
\author{Purusattam Ray}
\email{ray@imsc.res.in}
\affiliation{The Institute of Mathematical Sciences, Taramani, CIT Campus, Chennai 600113}

\date{\today}

\begin{abstract}
We report a novel critical behavior in the breakdown of an equal load sharing fiber bundle model at a dispersion $\delta_c$ of the breaking threshold of the fibers. For $\delta < \delta_c$, 
there is a finite probability $P_b$, that rupturing of the weakest fiber leads to the failure of the entire system. For $\delta \geq \delta_c$, $P_b = 0$. At $\delta_c, P_b \sim L^{-\eta}$, with $\eta \approx 1/3$, where $L$ is the size of the system. As $\delta \rightarrow \delta_c$, the relaxation time $\tau$ diverges obeying the finite size scaling law: $\tau \sim L^{\beta}(|\delta-\delta_c| L^{\alpha})$ with $\alpha, \beta = 0.33 \pm 0.05$. At $\delta_c$, the system fails, at the critical load, in avalanches (of rupturing fibers) of all sizes $s$ following the distribution $P(s) \sim s^{-\kappa}$, with $\kappa = 0.50 \pm 0.01$. We relate this critical behavior to brittle to quasi-brittle transition.

\end{abstract}


\maketitle


Fiber bundle model has received lots of attention over the years, as it is one of the simplest model that captures various importatnt aspects of the complicated problem of fracture \onlinecite{RevModPhys82}. 
Avalanche dynamics of brittle fracture and of plastic deformation, stress concentration, extreme 
statistics and related system size effect are some telltale features of fracture which are reproduced 
in this model. The model provides general understanding and insight of the physics of broad range of 
threshold activated dynamical systems.     
Here, we report another, hitherto unexplored, aspect of the model: a novel critical behavior in terms 
of the fluctuation in the disorder in the system. The criticality is directly related to the brittle to 
quasi-brittle transition in the model and have practical relevance in the context of such transitons 
in engineering brittle materials.   

Our study is based on the equal load sharing fiber bundle model \onlinecite{RevModPhys82} which consists 
essentially of Hookean fibers attached between two parallel bars. The bars are pulled apart with a force 
which induces a stress on the fibers. Each fiber sustains a stress up to a threshold (chosen randomly 
from a distribution) beyond which it breaks irreversibly. Once a fiber breaks, the stress of the 
fiber is redistributed equally among all the surviving fibers. This enhances the stress 
on the fibers and can cause further failure of some fibers and the rupturing process 
continues. Otherwise, the system attains a stable state with 
few broken fibers, when further increment of external load is required to make the fracture evolve. 
The evolution of fracture takes place in series of avalanches (burst) of rupturing of fibers as 
the stress is increased. The stress at which failure of the entire system happens is termed as the 
critical stress or strength of the system. 

In this letter, we study the fracture process in the model by tuning the width $\delta$ of the distribution 
$P(x)$ of the rupture threshold of the fibers. At low values of $\delta$, the failure of the system occurs, typically, 
starting from the rupturing of a single fiber, the fiber with minimum breaking threshold. Rupturing of 
one fiber triggers rupturing of several other fibers and the process continues till all the fibers are 
ruptured and the system fails. On the other hand, for large values of $\delta$, failure occurs through 
series of stable states at different stress levels. As the applied load is raised, the system goes through the 
states in succession of  avalanches of rupturing of fibers till the critical stress $\sigma_c$, when the remaining fraction $n_c$ of fibers ruptures bringing failure of the entire system. 
We ask the following questions, (i) is there a critical value $\delta_c$ demarcating these two regimes, 
(ii) what are the features of the transition at $\delta_c$ and (iii) what are the manifestations 
of the transition on the properties of the model and its relevance to the fracture properties in 
engineeering materials?

Here, we present results for uniform distribution of $P(x)$ over the window from $a$ to $(a+2\delta)$ 
within the range [0,1]. For uniform distribution, the possibility of a $\delta_c$ has been mentioned before 
\cite{Sornette}. We show, analytically, for the above distribution, $\delta_c = a/2$. For $\delta \leq\delta_c$, there is a finite probability $P_b$ that rupturing of the weakest fiber leads to the failure of the entire system.
$P_b \rightarrow 0$ as $\delta \rightarrow \delta_c$. At $\delta_c, P_b \sim L^{-\eta}$, with $\eta \approx 1/3$, where $L$ is the size of the system. We further show that, as $\delta$ approaches $\delta_c$, the relaxation time $\tau$ 
 diverges as $\tau \sim |\delta-\delta_c|^{-1}$ and the evolution of fracture becomes extremely slow. At $\delta_c$, the failure of the system occurs in succession of avalanches 
of urpturing of fibers. The avalanche size distribution $P(s)$ shows a scale free behavior and 
follows $P(s) \sim s^{-\kappa}$ with $\kappa \approx 1/2$.

At an applied stress $\sigma_{ext}$, the fraction of unbroken bond $n_u$ satisfies,
\begin{align}\label{eq:equation0}
1-n_u=\displaystyle\int_{0}^{\sigma_r}P(x)dx,
\end{align}
where $\sigma_r=\sigma_{ext}/n_u$ is the redistributed stress over the intact fibers. This gives, 
\begin{align}\label{eq:equation1}
n_u=\frac{1}{2}\left(1+\displaystyle\frac{a}{2\delta}\right)\pm\frac{1}{2}\left[\left(1+\displaystyle\frac{a}{2\delta}\right)^2-\displaystyle\frac{4\sigma_{ext}}{2\delta}\right]^{1/2}. 
\end{align}
At the failure point above equation has only one solution. This gives the critical stress $\sigma_c$ and critical fraction $n_c$ of the fibers at the failure point :
\begin{align}\label{eq:equation2}
\sigma_c=\frac{2\delta}{4}\left(1+\displaystyle\frac{a}{2\delta}\right)^2 \ \ \  \text{and} \ \ \ n_c=\frac{1}{2}\left(1+\displaystyle\frac{a}{2\delta}\right).
\end{align}
For $\delta \le \delta_c$ the fracture is abrupt ($n_c=1$). This gives, 
\begin{align}\label{eq:equation3}
\delta_c=a/2.
\end{align}

We now prove the critical divergence of $\tau$ as $\delta \rightarrow \delta_c$. From the dynamics 
of the system,
 the fraction $n_{t+1}$ of unbroken fibers at time step $(t+1)$ is expressed in terms of  the fraction $n_t$ 
at time $t$ as:
\begin{align}\label{eq:equation4}
n_{t+1}(\sigma_{ext})=\displaystyle\frac{1}{(2\delta)}\left[a+2\delta-\displaystyle\frac{\sigma_{ext}}{n_{t}(\sigma_{ext})}\right]. 
\end{align} 
At $\delta_c$, $\sigma_{ext} = a$. We solve the above equation for a slight deviation of $n_t(\sigma)$ from the critical point $n_c \ (=n_t(\sigma_c))$ by an amount $\Delta n$ such that : $n_t(\sigma)=n_c +\Delta n \ (\Delta n \rightarrow 0)$. Solving the above equation we get,
\begin{align}\label{eq:equation5}
\Delta n=Ae^{-t/\tau},
\end{align} 
where $A$ is a constant and $\tau \sim (\delta-\delta_c)^{-1}$.

To get better insight of the nature of the transition, we have studied the transition numerically. For 
simplicity, we have taken an uniform distribution for $P(x)$ of half width $\delta$ and mean at 0.5. For this distribution $\delta_c = 1/6$ (eqn.\eqref{eq:equation3}). Numerical results are obtained on $10^4$ configuration averages and for system sizes ranging from $10^2$ to $10^6$.
\begin{figure}[ht]
\centering
\captionsetup{justification=raggedright}
\includegraphics[width=9cm, keepaspectratio]{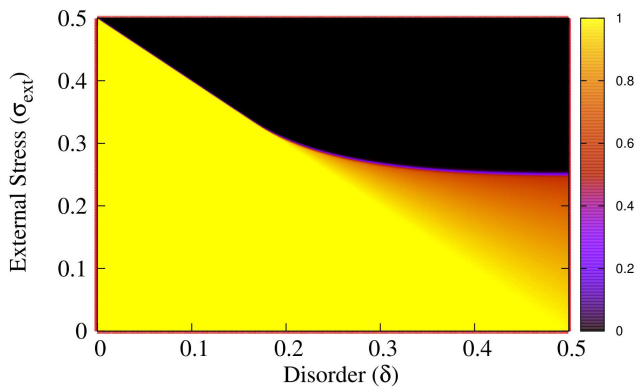}
\caption{(Color online) Contour plot for the response of the model to external stress. Black corresponds 
to the system with all the fibers broken ($n_u=0$).  Yellow corresponds to the initial condition where all fibers are intact ($n_u=1$). The color gradient is the partially broken phase ($0<n_u<1$).}
\label{fig:Phase_Diagram}
\end{figure}
Figure  \ref{fig:Phase_Diagram}  shows $n_u$ for different values of 
applied stress $\sigma_{ext}$ and $\delta$. Here, the black region denotes the system after complete 
failure ($n_u=0$) and yellow region represents the initial condition where all fibers are intact ($n_u=1$). The color gradient is the partially broken phase ($0<n_u<1$).  $\sigma_{ext}$ at the boundary of the black and yellow region denotes the critical stress $\sigma_c$. For $\delta<\delta_c$,  
$\sigma_c$  follows a straight line given by $a=0.5-\delta$. This is the region where we go from the 
yellow to black region by a single jump. 
For $\delta>\delta_c$ the critical stress deviates from the straight line and approaches to $\sigma_c=0.25$ at $\delta=0.5$ (\onlinecite{RevModPhys82}). In this region, fracture evolves 
through a series of partially broken stable phases at different levels of applied load.

%

\begin{figure}[ht]
\centering
\captionsetup{justification=raggedright}
\includegraphics[width=8cm, keepaspectratio]{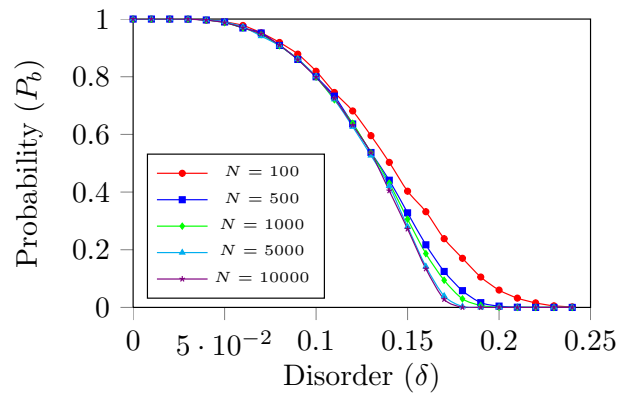}
\caption{Plot of $P_b$ with ($\delta$) for system sizes $L=100$ (oplus), $500$ (square), $1000$ (diamond), $5000$ (triangle) and $10000$ (star).}
\label{fig:Probability}
\end{figure}

Due to finite size of the system, even below $\delta_c$, global failure (rupture of all the fibers spanning 
the entire system) is not always obtained starting from the rupture of the weakest fiber. We determine 
$P_b$, the probability of system spanning fracture starting from the rupture of the weakest 
fiber, for various $\delta$ values and for various system sizes (see fig.\ref{fig:Probability}).
For a particular system size $L$, we define the critical $\delta$-value $\delta_c(L)$ at which $P_b$ 
goes to zero. It is quite clear that $\delta_c(L)$ has a system size dependence and approaches to 
$\delta_c = \delta_c(\infty)$ as $L \rightarrow \infty$. We find that $\delta_c(L)$ satisfies the finite-size 
scaling relation:   
\begin{align}\label{eq:equation6}
\delta_c(L)=\delta_c(\infty)+bL^{-1/\nu}, 
\end{align}
where the exponent $1/\nu$ has a value $0.33\pm0.02$ and $b$ is a constant. The error bar is determined
from the least square data fitting.

\begin{figure}[ht]
\centering
\captionsetup{justification=raggedright}
\includegraphics[width=8cm, keepaspectratio]{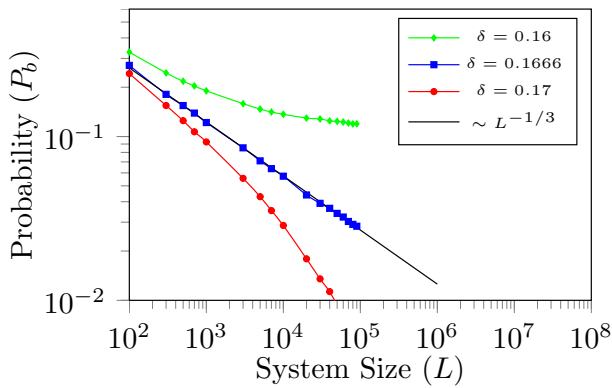}
\caption{Variation of $P_b$ with system size for $\delta=0.15$ (diamond), $0.1666$ (square) and $0.18$ (oplus). As a reference the plot with power law exponent $\eta\approx0.33$ (black solid line) is also given.}
\label{fig:Prob_System}
\end{figure}

Fig.\ref{fig:Prob_System} presents the variation of $P_b$ with system sizes in the neighborhood of 
$\delta_c$.
For $\delta<\delta_c$, $P_b$ tends to saturate and there is always a non-zero probability of rupturing 
of all the fibers starting from the single most vulnerable one. On the other hand, for 
$\delta>\delta_c$, $P_b$ falls off sharply to zero and probability of having an abrupt fracture 
vanishes in the thermodynamic limit ($L\rightarrow\infty$). At critical 
disorder $P_b$ shows a scale free behavior: 
\begin{align}\label{eq:equation7}
P(b) \sim L^{-\eta}
\end{align}
with an exponent $\eta=0.33\pm0.01$. The growth of fracture  
at $\delta_c$ is similar to the growth of incipient infinite cluster in various other statistical mechanical 
models like the growth of percolaing cluster at percolation threshold \cite{stauffer}.

\begin{figure}[ht]
\centering
\captionsetup{justification=raggedright}
\includegraphics[width=8cm, keepaspectratio]{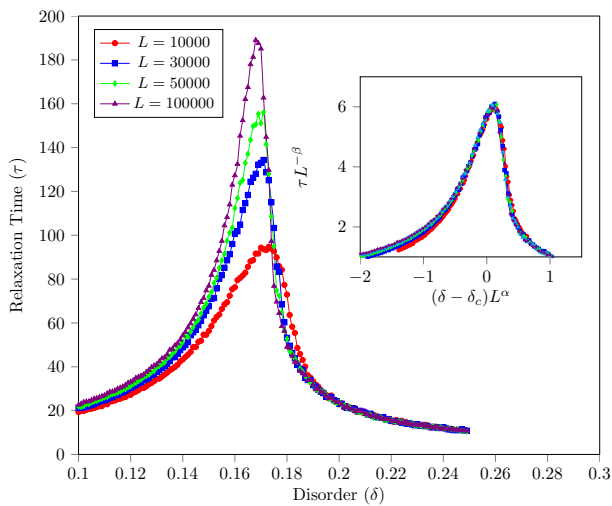}
\caption{Plot of relaxation time $\tau$ with $\delta$ for system sizes $L=10000$ (oplus), $30000$ (square), $50000$ (diamond) and $100000$ (triangle). At the inset $\tau L^{-\beta}$ 
is plotted against $(\delta-\delta_c)L^{1/\nu}$.}
\label{fig:Relaxation}
\end{figure}

To determine $\tau$, we apply the minimum load that is needed to rupture the weakest fiber and 
the system is allowed to evolve (keeping the load constant) till it fails or reaches a 
stable state with partially broken bonds.
$\tau$ is determined as the number of times the load is to be distributed among the unbroken fibers 
over the evolution of the system. 
Fig.\ref{fig:Relaxation} shows the variation of relaxation time $\tau$ with $\delta$ for different system 
sizes.
$\tau$ shows a maximum at $\delta=\delta_c(L)$ for system size $L$. As $L$ increases, the maximum value 
increases and the increament becomes sharper and sharper. Below $\delta_c$ the system experiences an abrupt failure within a few redistributing time steps starting from the rupture of the weakest fiber.
On the other hand, for $\delta > \delta_c$, due to a wide range of threshold values, the minimum stress 
corresponing to the weakest fiber is not sufficient to break all the fibers in the system. Staring from 
some intitial rupturing of the fibers, the system comes a stable configuration after few redistribution step for the 
stress. In both these cases, $\tau$ is finite. 

At $\delta=\delta_c$, the divergence of $\tau$ with the system size is found to obey the finite size 
scaling form: 
\begin{align}\label{eq:equation8}
\tau \sim L^{\beta}\Phi\big((\delta-\delta_c)L^{1/\nu}\big),
\end{align}
where $\beta=0.33\pm0.02$. Here, we have used the exponent $\nu$ from the perspective of a 
correlation length exponent. The scaling function $\Phi(x)$ assumes constant value for $x=0$ and 
for large $x$. Accordingly, at $\delta_c$, $\tau \sim L^{\beta}$, and close to $\delta_c$,
\begin{align}\label{eq:equation9}
\tau \sim (\delta-\delta_c)^{-\gamma}, \ \ \ \text{where} \ \ \ \gamma=\beta/\alpha=1.
\end{align}
The scaled $\tau$ is shown in the inset of fig.\ref{fig:Relaxation}. The scaling is sensitive to the choice of 
the value of the exponent $\beta$, which determines the error bar of it by simply checking the collapse 
in the scaling.



\begin{figure}[ht]
\centering
\captionsetup{justification=raggedright}
\includegraphics[width=8cm, keepaspectratio]{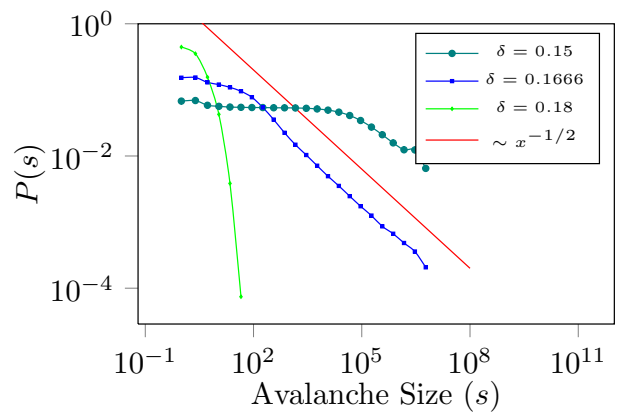}
\caption{Distribution of size of avalanches ($s$) prior to the application of minimum stress for disorder value in the neighborhood of $\delta_c$, typically $\delta=0.15$ (oplus), $0.1666$ (square) and $0.18$ (diamond). Reference line with power law exponent $\kappa=0.5\pm0.01$ is also given.}
\label{fig:Avalunche}
\end{figure}

Figure \ref{fig:Avalunche} shows the avalanche size distribution $P(s)$ verses the avalanche sizes $s$.
We start from applying a low load on the system, sufficient enough to rupture only the weakest fiber. 
Once a fiber is ruptered, the stress is redistributed among the remaining unbroken fibers. The breaking 
of fibers between any two consecutive stress redistribution steps  consitute an avalanche and the 
number of fibers broken 
gives the size of the avalanche. 
For $\delta < \delta_c$, the distribution tends to level off (upto a certain avalanche size) showing big avalanches are equally probable. In this region, after few big avalanches, the system fails completely 
within few time steps.  
In the region $\delta>\delta_c$, the distribution falls off very fast showing that, in this region, the 
system ceases to evolve after few small avalanches starting from the rupturing of the weakest fiber.  
At $\delta=\delta_c$ the distribution is a power law: 
\begin{align}\label{eq:equation10}
P(s) \sim s^{-\kappa},
\end{align}
where the exponent is given as $\kappa=0.5\pm0.01$.

In equal load sharing fiber bundle model, we see yet a novel critical behaviour at a 
critical disorder $\delta_c$ in the threshold strength of the individual fibers. The exponent values suggest 
that the critical behavior is in a different universality class than the behavior the model shows at the 
fracture point at the critical load $\sigma_c$ for $\delta > \delta_c$. For $\delta < \delta_c$, the model breaks abruptly. 
If each fiber breaks in a brittle manner, the overall fracture behavior of the entire system 
will also be brittle. For $\delta >\delta_c$, The fracturing of the system occurs througha series of 
partially broken stable phases which brings nonlinearity in the stress-strain behavior of the system and  
we get quasi-brittle like continuous fracture. $\delta_c$, in the model, corresponds to the brittle to 
quasi-brittle transition point.

Fracture in quasi-brittle materials has tremendous technological relevance and and as a result has received lots of attention in the past \onlinecite{Dieter,bazant}. Brittle materials, like glass, ceramics, wood, do not show much deformation under stress and fracture abruptly \onlinecite{Lawn}. Here, fracture is determined primarily 
by the nucleation of a vulnerable crack that spans through the system causing total failure. 
In heterogeneous materials, the quasi-brittle fracturing occurs by microcracking of the medium. It is characterized by complex temporal and spatial organizations of microcracks. This produces intermittency 
in the damage events and localization \onlinecite{bazant}.

We find that the transition at $\delta_c$ is marked by the divergence of relaxation time and 
fracture by creep. The creep phenomena and the avalanches in creep have been observed at the 
brittle to ductile transition point. It will be worth investigating to what extent the transition, that is 
observed in the simple fiber bundle model, is related to the brittle to quasi-brittle or ductile transition 
in engineering materials.

\end{document}